\documentclass[%
 reprint,
 amsmath,amssymb,
 aps,
]{revtex4-2}

\usepackage{graphicx}% Include figure files
\usepackage{dcolumn}% Align table columns on decimal point
\usepackage{bm}% bold math
\usepackage{amssymb}
\usepackage{amsmath,amssymb,mathtools} % in your preamble
\usepackage{placeins}
\usepackage{xcolor}
\usepackage[normalem]{ulem}
\usepackage{siunitx} % for \SI and \SIrange commands
\usepackage{tikz}
\usetikzlibrary{backgrounds}
\usetikzlibrary{arrows.meta,calc,positioning,decorations.pathreplacing}
\usepackage[colorlinks=true,linkcolor=blue,citecolor=blue,urlcolor=blue]{hyperref}

\begin{document}

\preprint{APS/123-QED}

\title{Stability analysis and quantum-limited noise properties of the Soliton-similariton fiber laser}% Force line breaks with \\

\author{Mohammad Iqbal Ashraf}
 \affiliation{Pratibimb Photonics Lab, School of Computing and Electrical Engineering, Indian Institute Of Technology, Mandi, Himachal Pradesh - 175005, India.}%Lines break automatically or can be forced with \\
 \author{Sreelakshmi Manjunath}
 \affiliation{Modelling and Intelligent Control Lab, School of Computing and Electrical Engineering, Indian Institute of Technology Mandi, Himachal Pradesh - 175005, India.}
\author{Srikanth Sugavanam}%

 \email{ssrikanth@iitmandi.ac.in}
%\affiliation{%
\affiliation{Pratibimb Photonics Lab, School of Computing and Electrical Engineering, Indian Institute Of Technology Mandi, Himachal Pradesh - 175005, India}%


%\affiliation{
%\affiliation{
 %Third institution, the second for Charlie Author
%\affiliation{%



\begin{abstract}
Soliton-similariton fiber lasers have demonstrated exceptional operational stability, maintaining continuous mode-locking for weeks despite large intracavity spectral and temporal breathing. We present the first stability study of this laser, rigorously establishing that the anomalous dispersion segment that supports the soliton is the cause of this robustness. Specifically, we perform linear stability analysis of the laser employing a Jacobian-based eigenvalue decomposition and show that the eigenvalues lie within the unit circle, leading to a positive stability margin, which is indicative of the robustness of the laser against small perturbations. Furthermore, the stability margin is observed to increase with the length of the anomalous fiber segment, clearly establishing its role in pulse stabilization. Critically, integrated pulse timing jitter and relative intensity noise as obtained from quantum noise-limited laser simulations are shown to be anti-correlated to the stability margin, further validating the results of the Jacobian analysis and establishing an unequivocal link between the reported noise performance of the soliton-similariton laser to the underlying pulse stabilization mechanism mediated by the anomalous segment. The direct link of the linear stability analysis to the underlying nonlinear physics of the laser, together with its significantly lower computational overhead, establishes it as a highly effective predictive framework for assessing laser noise performance, enabling novel approaches for designing quantum noise-limited ultrafast sources.
\end{abstract}
%\keywords{Rare-earth doped fiber amplifier, Soliton, Soliton propagation, Ultrashort pulse phenomena, Numerical simulations, Fiber laser, Fiber amplifier } %Use showkeys class option if keyword
\maketitle

%\tableofcontents

%\section{\label{sec:level1}First-level heading:\protect\\ The line
%break was forced \lowercase{via} \textbackslash\textbackslash}

\section{Introduction}

Ultrafast fiber lasers are indispensable tools in precision metrology, biomedical imaging, spectroscopy, and frequency comb technology. Their continued impact depends on designs that simultaneously deliver high pulse energies, short durations, and low noise. Early fiber lasers, developed from the late 1980s to the early 1990s, relied on creating solitons by balancing optical effects, but these were fundamentally limited to low energies \cite{HausIslam1985JQE,Kelly1992EL} and suffered from timing jitter \cite{HausMecozzi1993JQE,NamikiYuHaus1996JOSAB}. An improvement came with stretched-pulse lasers, which used alternating fiber types to allow the pulse to breathe, achieving nanojoule energies and better noise characteristics \cite{Tamura1993OL,Tamura1994APL,Namiki1997JQE_NoiseSPFL_I,Song2011OL_JitterRegimes}. However, this method also hit a scaling wall because at even higher energies, increased nonlinear phase leads to pulse distortion and multi-pulsing \cite{Tamura1995,Haus1995StretchedPulseAPM,Tang2005PRA}. The real breakthrough came with the discovery that parabolic-shaped pulses could be amplified to high energies without breaking apart via a process called self-similar propagation \cite{Anderson1993_JOSAB_WBFree}. Researchers like Fermann \emph{et al.} demonstrated that these amplifier similaritons provided a clear path to high-energy scaling, a concept that was quickly adapted into new laser designs \cite{Fermann2000_PRL_SelfSimilarPRL, Ilday2004_PRL_SelfSimilarLaser, Renninger2010_PRA_ANDiSelfSimilar}. This culminated in the development of the all-normal-dispersion laser, where a careful balance of spectral filtering, nonlinearity, dispersion, and gain shapes the pulse, creating a distinctive M-shaped light spectrum \cite{Chong2007_OL_ANDi20nJ, Chong2015_RPP_SimilaritonReview}. This self-similar evolution established a powerful new paradigm for generating high-energy pulses, fundamentally moving beyond the constraints of the older soliton and stretched-pulse regimes.

A particularly elegant extension emerged with hybrid soliton–similariton architectures. Oktem \emph{et al.} \cite{Oktem2010_NatPhoton_SSFL} demonstrated a cavity that combines a normal-dispersion gain fiber, generating parabolic similaritons, with an anomalous-dispersion passive fiber supporting soliton dynamics, separated by strong spectral filtering. Its elegance lies in two aspects. First, it does not attempt to manage or compensate for nonlinearity, but rather uses it to support two distinct classes of localized nonlinear optical structures within the same cavity. Second, this was the first experimental demonstration of amplifier similariton in a fiber laser cavity, where spectral filtering was employed for self-consistent seeding of the amplifier section of the cavity. This configuration exhibited exceptional stability with intensity noise at the $10^{-4}$ level over a radio-frequency range of 1–250 kHz, timing jitter of 27 fs measured from 1 kHz to the Nyquist limit, and weeks of continuous mode-locking without requiring careful noise engineering \cite{Oktem2010_NatPhoton_SSFL}. Subsequent work expanded the understanding of these systems through various lenses: birefringent filtering for all-fiber implementations \cite{Zhang2010_OL_SSFL_BirefringentFilter}, theoretical models of filter-enhanced stability \cite{Bale2010_OL_StrongSpectralFiltering}, transitions between operating states \cite{Wang2017_JOSAB_SpectralFilteringTransition}, real-time observation of instabilities and soliton molecules \cite{Lapre2019_SciRep_RealtimeInstabilities}, and comprehensive numerical simulations \cite{meng2020instabilities}. Recently, Mohamed \emph{et al.} \cite{Mohamed2024_NatCommun_EMS} proposed energy-managed soliton lasers as all-anomalous variants inspired by this architecture.
However, none of the works reported insofar \cite{Zhang2010_OL_SSFL_BirefringentFilter,Bale2010_OL_StrongSpectralFiltering,Wang2017_JOSAB_SpectralFilteringTransition,Lapre2019_SciRep_RealtimeInstabilities,meng2020instabilities,Mohamed2024_NatCommun_EMS} attempt to explain the exceptional stability and the observed weeks of mode-locking of this configuration vis-à-vis other fiber laser cavities, given that the pulse experiences strong intra-cavity spectral and temporal breathing. This kind of stability is uncommon and consequential for ultralow-noise applications, such as frequency combs and optical sampling systems. 

In this work, we rigorously demonstrate that the soliton-supporting anomalous segment in the hybrid cavity is responsible for their observed robustness. The soliton-shaping in this segment, because of its inherent nature, generates a restoring force that actively suppresses external perturbations and filters internal noise, for example, amplified spontaneous emission (ASE) noise in the gain fiber. To substantiate this claim, we construct a second cavity by replacing the anomalous segment with a normally dispersive fiber and compare the two configurations using linear stability analysis and quantum-limited noise characterization, two frameworks that, as we demonstrate, provide complementary and mutually reinforcing evidence for the robustness of the hybrid cavity. The basin of attraction analysis confirms that both achieve mode-locking and reach a steady state from arbitrary initial conditions. However, linear stability analysis via Jacobian eigenvalue decomposition of their steady state reveals that the hybrid cavity exhibits a spectral radius less than one, demonstrating that it is asymptotically stable to external perturbations, whereas the all-normal variant shows a spectral radius greater than one, indicating that it is susceptible to external perturbations. Systematic variation of the anomalous fiber length further demonstrates that increasing anomalous contribution monotonically reduces the spectral radius or, equivalently, increases the stability margin. Quantum-limited noise characterization \cite{Paschotta2004_APB_NoiseI,Paschotta2004_APB_NoiseII} reveals the soliton-similariton configuration achieves timing jitter and relative intensity noise (RIN) orders of magnitude lower than the all-normal variant, demonstrating the filtering of ASE-induced noise generated in the gain fiber. Thus, the restoring force generated by the anomalous fiber accounts for the asymptotic stability and superior quantum-limited noise performance of soliton–similariton fiber lasers. In contrast, the absence of such a restoring eigenstate in the all-normal variant leads to instability and degraded noise performance.  Furthermore, a systematic evaluation of the stability margin and integrated timing jitter across various anomalous fiber lengths in the soliton-similariton configuration reveals a strong anticorrelation between these metrics. This observation motivates the introduction of stability margin as a general predictive metric for fiber laser optimization, serving the same purpose as integrated timing jitter while requiring orders of magnitude fewer simulation iterations. This not only reconciles the long-standing observations of exceptionally low noise in this hybrid cavity but also establishes a computationally efficient framework for designing and optimizing fiber lasers in general \cite{Oktem2010_NatPhoton_SSFL,Grelu2012_NatPhoton_DissipativeSolitons}.

The paper is organized as follows. Section II outlines the cavity schematics, methodology, and validation. Section III analyzes both the hybrid and all-normal cavities via Poincaré maps and confirms the existence of steady states. Section IV presents the linear-stability results and attributes the hybrid cavity’s superior stability to the anomalous segment. Section V reports the quantum-limited noise characterization, demonstrating the superior noise performance of the soliton-similariton laser and its strong correlation with the stability margin obtained from the linear stability analysis.

\section{Methodology}

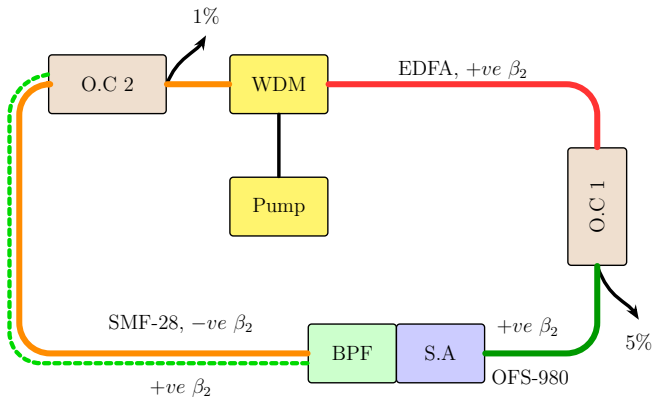
\begin{figure}[t]
\centering
\resizebox{\columnwidth}{!}{%
\begin{tikzpicture}[
    font=\large,
    line cap=round,
    line join=round,
    box/.style={
        draw, thick, rounded corners=2pt,
        minimum width=2.4cm, minimum height=1.2cm,
        align=center, fill=brown!25
    },
    boxV/.style={
        draw, thick, rounded corners=2pt,
        minimum width=1.2cm, minimum height=2.4cm,
        align=center, fill=brown!25
    },
    ybox/.style={
        draw, thick, rounded corners=2pt,
        minimum width=2.0cm, minimum height=1.2cm,
        align=center, fill=yellow!70
    },
    gbox/.style={
        draw, thick, rounded corners=2pt,
        minimum width=1.8cm, minimum height=1.2cm,
        align=center, fill=green!20
    },
    bbox/.style={
        draw, thick, rounded corners=2pt,
        minimum width=1.8cm, minimum height=1.2cm,
        align=center, fill=blue!20
    },
    fiberR/.style={line width=3.5pt, draw=red!80},
    fiberG/.style={line width=3.5pt, draw=green!60!black},
    fiberY/.style={line width=3.5pt, draw=orange!90!yellow},
    tap/.style={line width=2pt, draw=black, -{Stealth[length=3mm]}},
]

% =========================
% Components
% =========================
\node[box]  (C2)   at (0,4)     {O.C 2};
\node[ybox] (WDM)  at (3.5,4)   {WDM};
\node[ybox] (Pump) at (3.5,1.5) {Pump};
\node[boxV] (C1) at (10.0,1.5) {\rotatebox{90}{O.C 1}};
\node[gbox] (BPF) at (5.0,-1.5) {BPF};
\node[bbox] (SA)  at (6.8,-1.5) {S.A};

% =========================
% Coordinates
% =========================
\coordinate (TopL) at (-1.8,4);
\coordinate (BotL) at (-1.8,-1.5);
\coordinate (BotR) at (10.0,-1.5);
\coordinate (TopR) at (10.0,4);

\coordinate (C1TopMid) at ($(C1.north west)!0.5!(C1.north east)$);
\coordinate (C1BotMid) at ($(C1.south west)!0.5!(C1.south east)$);

% =========================
% Yellow fiber
% =========================
\draw[fiberY, rounded corners=18pt]
    (C2.west) -- (TopL) -- (BotL) -- (BPF.west);

% =========================
% Dashed beta2 section
% =========================
\coordinate (TopL2) at (-2.0,4);
\coordinate (BotL2) at (-2.0,-1.7);
\draw[line width=2.5pt, dashed, draw=green!85!black, rounded corners=18pt]
    ($(C2.west)+(0,0.2)$) -- (TopL2) -- (BotL2) -- ($(BPF.west)+(0,-0.2)$);

% =========================
% O.C 2 tap
% =========================
\begin{scope}[on background layer]
\draw[tap]
    (C2.east) .. controls +(0.2,0.55) and +(-0.2,-0.35)
    .. ($(C2.east)+(0.8,1.0)$);
\node at ($(C2.east)+(0.8,1.4)$) {1\%};
\end{scope}

\draw[fiberY] (C2.east) -- (WDM.west);
\draw[line width=2pt] (Pump.north) -- (WDM.south);

% =========================
% Labels
% =========================
\path (WDM.east) -- (TopR)
node[midway, above] {EDFA, $+ve~\beta_2$};

\node at (1.5,-2.2) {$+ve~\beta_2$};
\node at (1.5,-0.9) {SMF-28, $-ve~\beta_2$};

% =========================
% Red fiber
% =========================
\draw[fiberR, rounded corners=18pt]
    (WDM.east) -- (TopR) -- (C1TopMid);

% =========================
% O.C 1 tap
% =========================
\begin{scope}[on background layer]
\draw[tap]
    (C1BotMid) .. controls +(0.25,-0.7) and +(-0.25,0.4)
    .. ($(C1BotMid)+(0.85,-1.1)$);
\node at ($(C1BotMid)+(0.85,-1.55)$) {5\%};
\end{scope}

% =========================
% GREEN fiber — labels aligned, opposite sides
% =========================
\draw[fiberG, rounded corners=18pt]
    (C1BotMid) -- (BotR) -- (SA.east)
    node[pos=0.62, above=5pt] {$+ve~\beta_2$}
    node[pos=0.60, below=5pt] {OFS-980};

\end{tikzpicture}%
}
\caption{\textbf{Schematic of the soliton-similariton fiber laser cavity.} Pulse propagates clockwise through erbium-doped fiber amplifier (EDFA, red fiber, $+\beta_2$), Coupler 1, OFS-980 fiber (green, $+\beta_2$), bandpass filter and saturable absorber (BPF/S.A., green/blue box), SMF-28 fiber (orange, $-\beta_2$), and Coupler 2. Output couplers (orange) extract light from the cavity: Coupler 1 with 5\% coupling ratio provides the main output, while Coupler 2 with 1\% coupling ratio serves as a monitoring tap. The pump laser is coupled via wavelength-division multiplexer (WDM). The schematic of the all-normal-dispersion fiber laser is obtained by replacing the SMF-28 passive fiber (orange line) with a positive-dispersion fiber (green dashed line).}
\label{fig:laser_schematic}
\end{figure}

FIG.~\ref{fig:laser_schematic} shows the schematics of the soliton-similariton fiber laser cavity used in the proceeding analysis. The cavity consists of an erbium-doped fiber (EDF) that provides distributed, bandwidth-limited saturable gain, a \(5\%\) output coupler, a normally dispersive single-mode fiber (OFS-980) segment, a lumped saturable absorber (SA), a narrowband bandpass filter centered at \(1550\,\mathrm{nm}\), an anomalous single-mode fiber (SMF-28), and a second \(1\%\) output coupler.

\begin{figure*}[ht]
  \centering
 
 \includegraphics[width=1.0\textwidth]{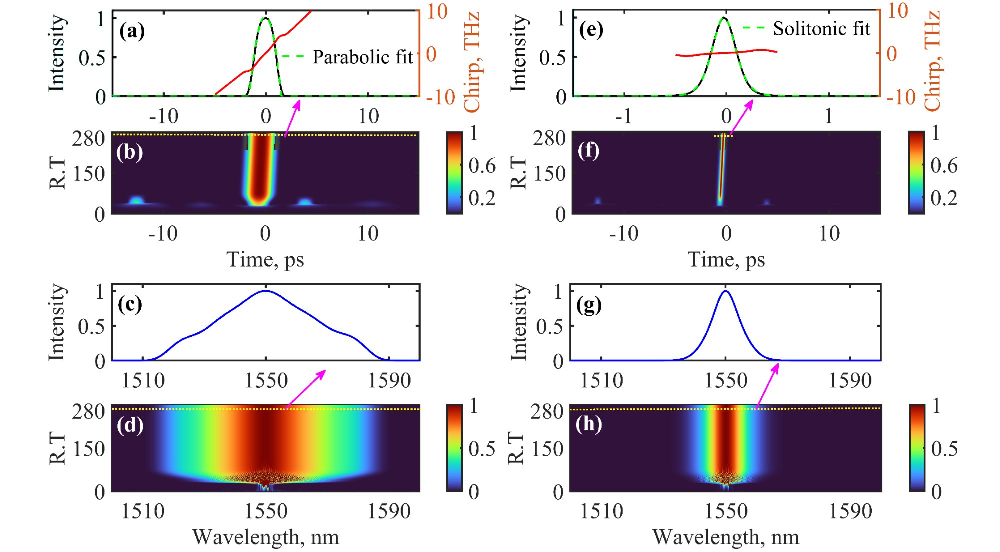}
  \caption{\textbf{Pulse formation dynamics and steady-state output at the two output ports of the soliton-similariton fiber laser.} \textbf{Left ($\mathrm{OC}_1$):} Similariton output after the normal-dispersion segment. \textbf{Right ($\mathrm{OC}_2$):} Soliton output after the anomalous segment. Panels (a,e): steady-state temporal intensity (black) with parabolic/solitonic fits (green dashed) and their respective chirp (red, right axis). Panels (b,f): temporal evolution over 300 round trips showing convergence from noise. Panels (c,g): steady-state spectra centered at $1550\,\mathrm{nm}$. Panels (d,h): spectral evolution over round trips. All intensities are normalized. $\mathrm{R.T}$ represents round trips}

  \label{fig:Steady state diagnostics}
\end{figure*}
\begin{figure}[ht!]
\centering
\includegraphics[width=1.05\linewidth]{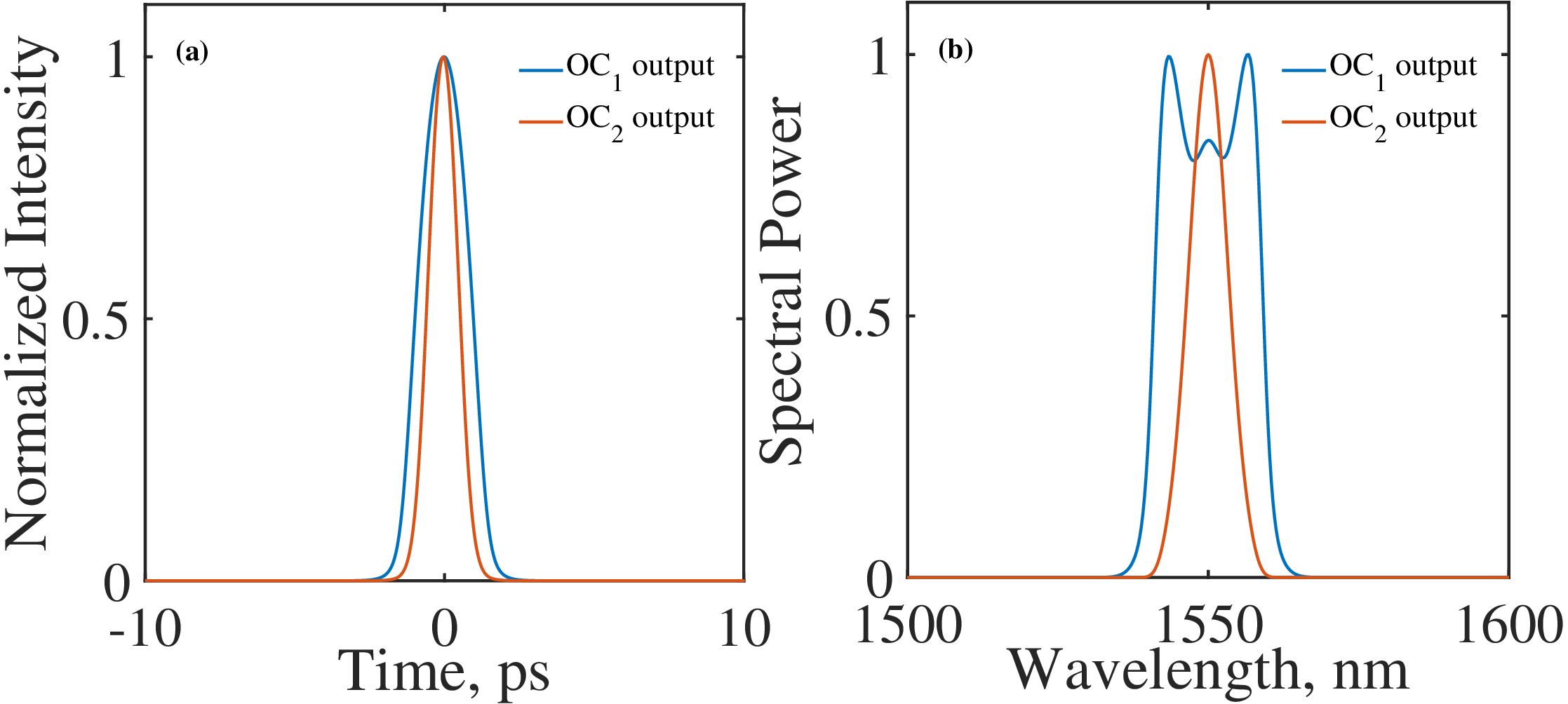}
\caption{\textbf{Steady-state output of the all-normal dispersion fiber laser.} The anomalous segment in Fig.~\ref{fig:laser_schematic} is replaced by a normal-dispersion fiber. The normalized temporal intensity and spectral power are shown after 300 round trips.}

\label{fig:Output pulse of the cavity varian}
\end{figure}

Field evolution in each fiber segment is modeled by the generalized nonlinear Schrödinger equation (GNLSE) \cite{Oktem2010_NatPhoton_SSFL}:
\begin{equation}
\begin{aligned}
\frac{\partial A}{\partial z}
&= -\,i\frac{\beta_2}{2}\frac{\partial^2 A}{\partial t^2}
  + \frac{\beta_3}{6}\frac{\partial^3 A}{\partial t^3}
  + \frac{g-\alpha}{2}\,A\\
&\quad + i\gamma |A|^2 A
  - i\,\gamma\,\tau_R\,A\,\frac{\partial}{\partial t}\!\big(|A|^2\big).
\end{aligned}
\label{eq:GNLSE}
\end{equation}
Here \(A(z,t)\) is the slowly varying envelope, \(\beta_2\) is group velocity dispersion (GVD), \(\beta_3\) is third-order dispersion coefficient, \(\gamma\) is the Kerr nonlinearity, \(\tau_R\) is the Raman response time, \(g\) is the gain corresponding to EDF, and \(g=0\) for passive fibers. The parameter $ \alpha$ represents the linear loss of the optical fibers. The gain $g$ is modeled as a spectrally limited Lorentzian \cite{Oktem2010_NatPhoton_SSFL,Yarutkina2013OE}, written in compact form as \(g(\omega)=g_{\mathrm{SS}}\!\big/\!\big(1+W/W_0+(\omega-\omega_0)^2/\Delta\omega^2\big)\), where \(g_{\mathrm{SS}}\) is the small-signal gain, \(W(z)=\int |A(z,t)|^2\,\mathrm{d}t\) is the pulse energy, \(W_0\) is the saturation energy, and \(\Delta\omega\) corresponds to an effective gain bandwidth  around centre frequency \(\omega_0\). The saturable absorber (SA) is modeled as a lumped, intensity-dependent transmission \cite{Oktem2010_NatPhoton_SSFL} using \(T(t)=1-q_0\!\big/\!\big(1+P(t)/P_0\big)\) with \(P(t)=|A(t)|^2\), where \(q_0\) is the modulation depth and \(P_0\) is the saturation power. 

Simulation parameters are similar to those used in \cite{Oktem2010_NatPhoton_SSFL}. The cavity comprises a 1~m erbium-doped fiber (EDF) with $\beta_2 = +76.9~\mathrm{ps}^2/\mathrm{km}$, $\beta_3 = 0.168~\mathrm{ps}^3/\mathrm{km}$, saturation power $P_{\mathrm{sat,G}} = 200~\mathrm{mW}$, small-signal gain $g_{\mathrm{SS}} = 3.35~\mathrm{m}^{-1}$, and gain bandwidth $\Delta\omega$ corresponding to $\sim$50~nm FWHM centered at 1550~nm. The passive fiber segments consist of a 0.65~m OFS-980 section ($\gamma = 2.1~\mathrm{W}^{-1}\mathrm{km}^{-1}$, $\beta_2 = 4.5~\mathrm{ps}^2/\mathrm{km}$, $\beta_3 = 0.109~\mathrm{ps}^3/\mathrm{km}$) and approximately 3~m of SMF-28 ($\gamma = 1.1~\mathrm{W}^{-1}\mathrm{km}^{-1}$, $\beta_2 = -22.8~\mathrm{ps}^2/\mathrm{km}$, $\beta_3 = 0.086~\mathrm{ps}^3/\mathrm{km}$). Passive fiber loss is $\alpha =- 0.2~\mathrm{dB}$. The saturable absorber is characterized by saturation power $P_0 = 2.13~\mathrm{kW}$ and modulation depth $q_0 \geq 0.7$, consistent with prior modeling of similar cavities \cite{Renninger2012JSTQE,meng2020instabilities}. The intracavity bandpass filter is implemented as a lumped Gaussian spectral transfer function with 10--15~nm FWHM centered at 1550~nm. The SMF-28 length determines the net cavity dispersion; all other component lengths remain fixed.

To isolate the role of the anomalous-dispersion segment in aiding the stability, we analyze two cavity configurations that differ only in the dispersion sign of the third passive fiber: the soliton-similariton laser (System~1) employs anomalous-dispersion fiber (SMF-28) with $\beta_{2} = -22.8~\mathrm{ps}^2/\mathrm{km}$, enabling dual nonlinear attraction mechanisms through both soliton and similariton dynamics. An all-normal variant (System~2) replaces this segment with a normal-dispersion fiber having $\beta_{2} = +22.8~\mathrm{ps}^2/\mathrm{km}$ of equal magnitude, eliminating soliton formation while preserving similariton shaping. All other cavity parameters, fiber lengths, nonlinearity, gain, saturable absorption, and filtering remain identical between the two systems, allowing direct attribution of any performance differences to the presence or absence of the anomalous segment.

Propagation through distributed segments uses a symmetric split-step Fourier method \cite{AGRAWAL817042}: dispersion and gain are applied in the frequency domain, Kerr and Raman terms in the time domain, and lumped elements (SA, filter, output couplers) act as instantaneous maps between segments. Numerical simulations employ  $N = 1024$ grid points over a temporal window $T_{\mathrm{window}} = 30~\mathrm{ps}$, providing time resolution $\Delta t \approx 29.30~\mathrm{fs}$ and spectral resolution $\Delta f = 33.3~\mathrm{GHz}$. We denote the gain fiber by \(\mathrm{GF}\), the normal-dispersion passive fiber by \(\mathrm{PF}_1\) (OFS-980), the anomalous-dispersion passive fiber by \(\mathrm{PF}_2\) (SMF-28), the saturable absorber by \(\mathrm{SA}\), the bandpass filter by \(\mathrm{BPF}\), and the two output couplers by \(\mathrm{OC}_1\) and \(\mathrm{OC}_2\). 

FIG.~\ref{fig:Steady state diagnostics} summarizes pulse build-up and steady-state outputs at the two couplers. The temporal and spectral build-up from quantum noise over 300 round trips are shown in panels (b,f) and (d,h), respectively, confirming rapid convergence to stable mode locking. The steady-state output at $\mathrm{OC}_1$ as shown in panels (a,c), after the normally dispersive gain fiber, is a linearly chirped parabolic similariton with a broad spectrum. The steady-state output at $\mathrm{OC}_2$, as shown in panels (e,g), after the anomalous segment, approaches a near-transform-limited soliton with minimal chirp and a narrower spectrum. This is in qualitative agreement with the reported experimental result \cite{Oktem2010_NatPhoton_SSFL}.

\section{Attractor Dynamics}

Soliton-similariton fiber lasers are known to exhibit remarkable operational stability, maintaining mode-locking for weeks without active stabilization~\cite{Oktem2010_NatPhoton_SSFL}. This robustness suggests convergence to a stable fixed point with a large basin of attraction. To visualize this more clearly, we compare the two cavity configurations mentioned in previous sections, which differ only in the dispersion sign of the third passive fiber (PF2). FIG.~\ref{fig:Output pulse of the cavity varian} shows the steady-state pulse profiles at OC1 and OC2 for the all-normal variant (System~2), confirming stable mode-locked operation. By examining how these cavities respond to diverse initial conditions, we can assess whether the anomalous segment plays a critical role in their convergence behavior.

To visualize the convergence dynamics, we construct a discrete Poincaré map by sampling the circulating field once per round trip $k$ at \(\mathrm{OC}_2\) and projecting it onto two low-order, physically interpretable and observable features:
\begin{equation}
\mathbf{x}_k=\bigl( T_{\mathrm{RMS}}(k),\,\omega_{\mathrm{RMS}}(k)\bigr),
\end{equation}
where the temporal and spectral RMS widths $\bigl( T_{\mathrm{RMS}}(k),\,\omega_{\mathrm{RMS}}(k)\bigr)$ are computed from the intensity $I(t)=|A(t)|^2$ and spectrum $S(\omega)=|\mathcal{F}\{A\}|^2$ of the intracavity field $A(t)$ at \(\mathrm{OC}_2\). This two-dimensional projection captures the essential pulse breathing dynamics.

To test the robustness of convergence, we initialize simulations from five maximally diverse quantum noise seeds while keeping the total input energy fixed. These seeds include (i) white noise, (ii) narrowband colored noise ($\Delta\omega\!\approx\!0.15$), (iii) wideband noise with a small positive carrier offset ($+\omega_0$), (iv) medium-narrow noise with negative offset ($-\omega_0$) plus slow amplitude modulation, and (v) higher-photon white noise multiplied by a broad burst envelope. These choices serve a simple purpose:  convergence of trajectories $\{\mathbf{x}_k\}$ from such maximally diverse initial conditions to the same terminal point $\mathbf{x}_\infty$, suggests a wide basin of attraction.

Convergence is monitored via the relative change in pulse energy or power between consecutive round trips $k$ and $k+1$ defined as \cite{schreiber2007study}:
\begin{equation}
\varepsilon_{k}^{\text{energy}} \;=\; \frac{\displaystyle\sum_{j=1}^{N} |A_{k+1}^{j}|^{2} - \sum_{j=1}^{N} |A_{k}^{j}|^{2}}{\displaystyle\sum_{j=1}^{N} |A_{k}^{j}|^{2}},
\label{eq:energy_convergence}
\end{equation}
where $A_{k}^{j}$ denotes the field amplitude at round trip $k$ and grid point $j$. Convergence is declared when $\varepsilon_{k}^{\text{energy}} < \varepsilon_{\text{tol}}$ for all $k \ge k_{\text{conv}}$, where,$\varepsilon_{\text{tol}}$ is the tolerance parameter and \(k_{\text{conv}}\) is the smallest round-trip index such that the convergence criterion is satisfied for all subsequent round trips. In this work, we use \(\varepsilon_{\text{tol}} = 10^{-6}\) for standard convergence and \(\varepsilon_{\text{tol}} = 10^{-9}\) when performing the stability analysis.
\begin{figure*}
  \centering
   \includegraphics[width=1.02\linewidth]{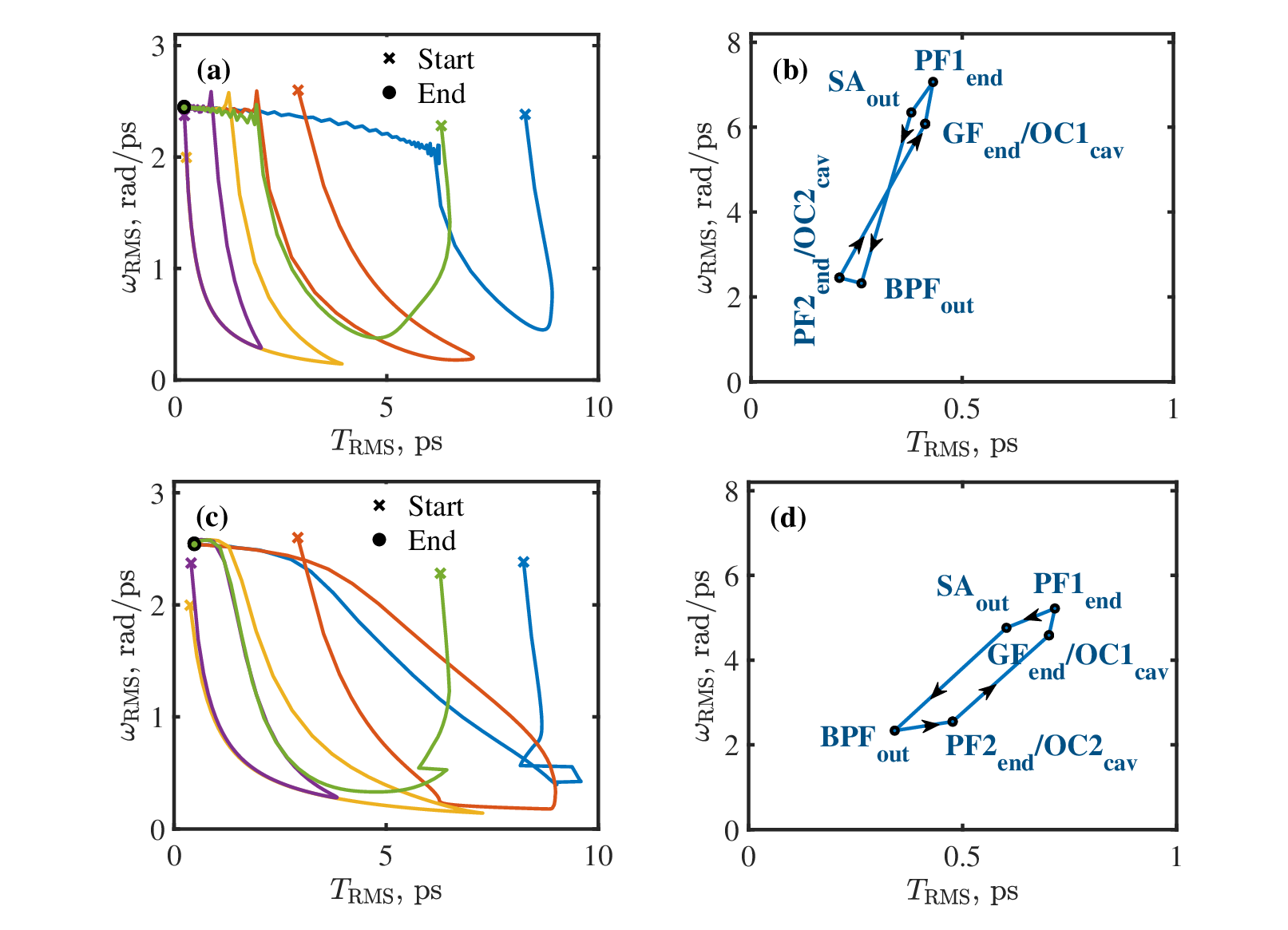} 
\caption{\textbf{Convergence dynamics and intracavity evolution for soliton-similariton and all-normal fiber lasers.} \textbf{(a,c)} Phase-space trajectories in the Poincaré plane $(T_{\mathrm{RMS}}, \omega_{\mathrm{RMS}})$ for five independent quantum-noise initial conditions (marked ×) converging to fixed points (filled dots) for \textbf{(a)} soliton-similariton and \textbf{(c)} all-normal configurations. Both cavities exhibit wide basins of attraction, but the convergence character differs. The soliton-similariton laser shows oscillatory relaxation due to competing soliton and similariton dynamics, while the all-normal variant converges smoothly via purely dissipative pulse shaping. The number of round-trips taken to reach steady state for each noise is slightly different; however, the plot shown is for a fixed round-trip of 250.  \textbf{(b,d)} Intracavity Poincaré maps reveal pulse evolution through successive cavity elements once the steady state has been reached. Closed trajectories confirm fixed-point operation in both systems, with arrows indicating propagation direction.}
\label{fig:fixed_point_attractor}
\end{figure*}
\begin{figure*}[ht]
    \centering
    \includegraphics[width=\textwidth]{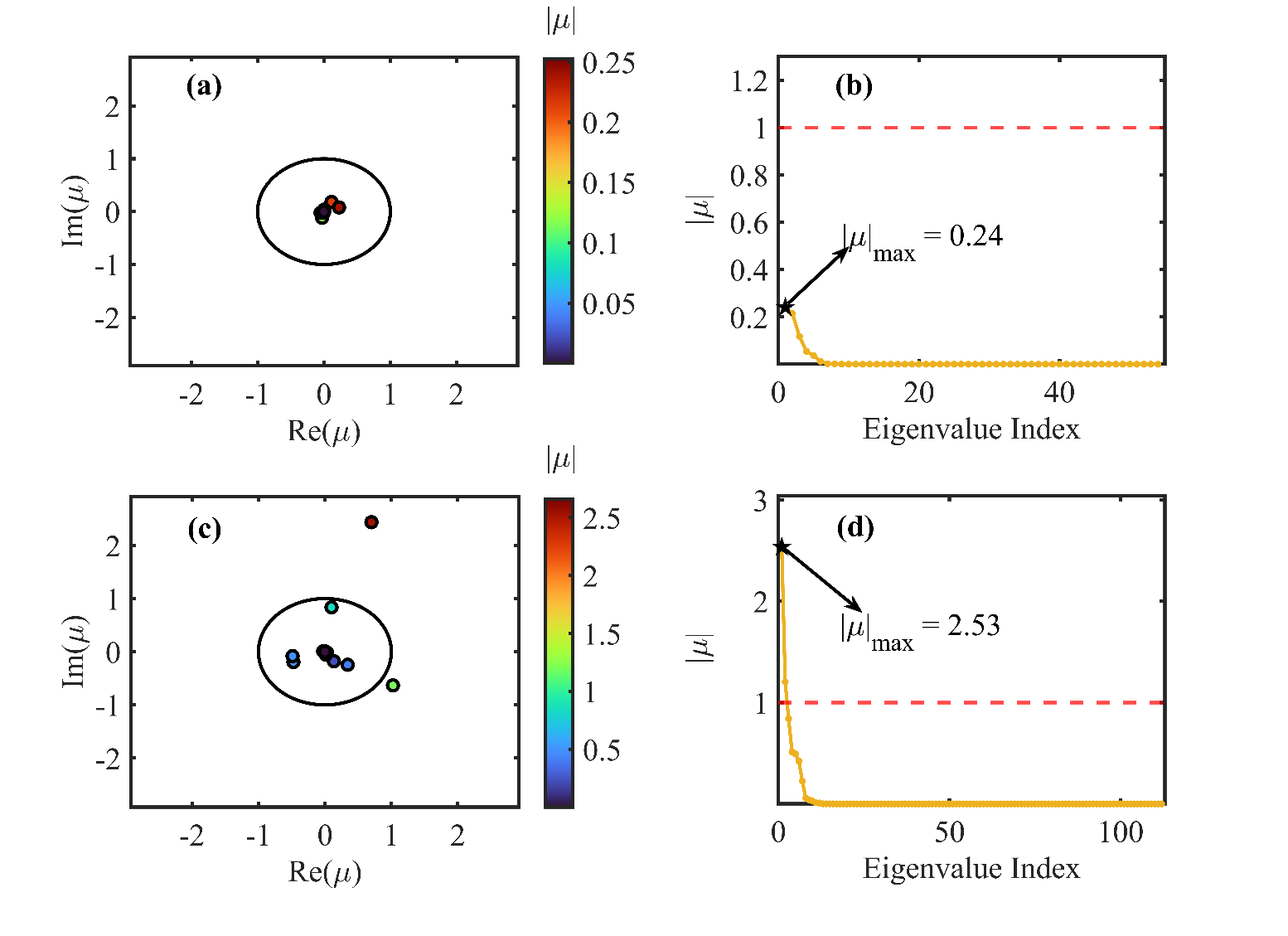}
    \caption{\textbf{Linear stability comparison.}
Top row: \emph{soliton-similariton laser}. Bottom row: \emph{All-normal fiber laser}.
\textbf{(a,c)} Eigenvalues $\mu$ of the round-trip linearization in the complex plane; the unit circle marks the stability boundary $|\mu|=1$. The colorbar encodes the absolute value of eigenvalues $|\mu_i|$.
\textbf{(b,d)} Sorted magnitudes $|\mu|$ with the red dashed line at $|\mu|=1$.
All eigenvalues of the soliton-similariton laser cavity lie inside the unit circle and exhibit a larger spectral gap($1-\rho_{\mathbf{J}}$), indicating greater robustness of the fixed point. In contrast, the all-normal variant design has several eigenvalues ($|\mu|>1$) outside the unit circle, indicating that the fixed point associated with it is susceptible to perturbation.}
\label{fig:stability_analysis}
\end{figure*}

FIG.~\ref{fig:fixed_point_attractor}(a,c) shows the phase-space trajectories in the Poincaré plane for both cavity configurations. In both cases, all five diverse initial conditions (marked with $\times$) converge toward a common terminal point (filled dot), suggesting that both cavities exhibit attractor-like behavior with wide basins of attraction. However, the character of convergence differs notably between the two designs. The soliton-similariton laser approaches its steady state with a noticeable overshoot and undershoot (FIG.~\ref{fig:fixed_point_attractor}(a)). This can be qualitatively explained as follows: the gain fiber pushes the pulse toward a parabolic (similariton) profile, while the anomalous-dispersion segment reshapes it toward a soliton. This push-pull competition produces small overshoots and undershoots before the system reaches its steady state. In contrast, the all-normal variant exhibits smooth, monotonic convergence (FIG.~\ref{fig:fixed_point_attractor}(c)) since only the parabolic shaping remains, and there are no competing dynamics.
FIG.~\ref{fig:fixed_point_attractor}(b,d) shows the intra-cavity Poincaré map for both cavities once the steady state has been reached. Markers trace the pulse state at successive element outputs with arrows indicating propagation direction and ``end'' denoting the output of each element. The map forms a closed polygon, indicating that the pulse state at the end of the gain fiber coincides with itself one round trip later, a graphical signature of a fixed point in the discrete cavity map.

These Poincaré maps reveal that both configurations converge from diverse initial conditions to fixed points with wide attraction basins but with markedly different convergence characteristics.  Conventionally, the oscillatory convergence of soliton-similariton would raise stability concerns, yet these lasers demonstrate weeks of stable mode-locking without active stabilization~\cite{Oktem2010_NatPhoton_SSFL}. To resolve this paradox, we perform linear stability analysis via Jacobian eigenvalue decomposition for both cavities.

\section{Linear Stability Analysis}
Linear stability analysis of mode-locked fiber lasers quantifies the stability of the steady-state fixed point by examining the eigenvalue spectrum of the linearized round-trip operator. Previous approaches have constructed this operator by analytically linearizing the governing equations-either the master equation in the small-variation limit~\cite{Kapitula2002} or the concatenated transfer functions of lumped cavity models~\cite{Shinglot2025}. The latter yields the monodromy operator whose spectrum decomposes into an analytically derivable essential spectrum and discrete eigenvalues associated with localized perturbation modes. This framework provides rigorous stability criteria but requires explicit linearization of each cavity element.

Here, we construct the linearized operator directly from the nonlinear dynamics. We perturb the steady-state field along each independent degree of freedom, propagate each perturbation through one full round trip, and assemble the Jacobian from the resulting response. This numerical differentiation approach treats one full round trip as a black box, bypassing analytical linearization while retaining full fidelity to the underlying physics, including saturable gain, self-phase modulation, and filtering, as they occur spatially throughout the cavity. We leverage this framework to systematically vary cavity parameters and track how stability evolves, revealing the physical mechanisms that stabilize the pulse.

The laser cavities can be treated as discrete-time dynamical systems in which the complex field envelope $A_k(t)$ at the beginning of the $k^{\mathrm{th}}$ round trip is transformed into the field of the next round trip by a round-trip operator $\mathcal{F}$. The evolution is written as
\begin{equation}
A_{k+1} = \mathcal{F}(A_k),
\end{equation}

The operator $\mathcal{F}$ represents the cumulative action of all cavity elements during one circulation. Using right-to-left composition,
\begin{equation}
\mathcal{F} =
\mathrm{OC}_{2}\circ\mathrm{PF}_{2}\circ\mathrm{BPF}\circ\mathrm{SA}\circ\mathrm{PF}_{1}\circ\mathrm{OC}_{1}\circ\mathrm{GF}
\end{equation}
A steady-state pulse $A^{\mathrm{st}}(t)$ is a fixed point satisfying $A^{\mathrm{st}} = \mathcal{F}(A^{\mathrm{st}})$. Henceforth, all fields are taken to be implicit functions of time $t$.

To examine stability under small perturbations, we consider a perturbed field:
\begin{equation}
A_k = A^{\mathrm{st}} + \delta A_k,
\end{equation}
where $\delta A_k$ represents a small perturbation from the fixed point at round trip $k$. Substituting into the evolution equation yields:
\begin{equation}
A^{\mathrm{st}} + \delta A_{k+1} = \mathcal{F}(A^{\mathrm{st}} + \delta A_k).
\end{equation}
Performing a first-order Taylor expansion of $\mathcal{F}$ around the fixed point:
\begin{equation}
\mathcal{F}(A^{\mathrm{st}} + \delta A_k) \approx \mathcal{F}(A^{\mathrm{st}}) + \frac{\partial \mathcal{F}}{\partial A}\bigg|_{A=A^{\mathrm{st}}} \cdot \delta A_k + \mathcal{O}(|\delta A_k|^2).
\end{equation}
Since $\mathcal{F}(A^{\mathrm{st}}) = A^{\mathrm{st}}$ by definition, and for perturbations satisfying $|\delta A_k| \ll |A^{\mathrm{st}}|$ where the quadratic and higher-order terms are negligible, the linearized perturbation dynamics become:
\begin{equation}
\delta A_{k+1} \approx \mathbf{J} \cdot \delta A_k,
\end{equation}
where $\mathbf{J} = \partial\mathcal{F}/\partial A|_{A=A^{\mathrm{st}}} \in \mathbb{C}^{N \times N}$ is the Jacobian matrix evaluated at the fixed point and $N$ are temporal grid points.
\begin{figure*}[ht!] 
\centering 
\includegraphics[width=\linewidth]{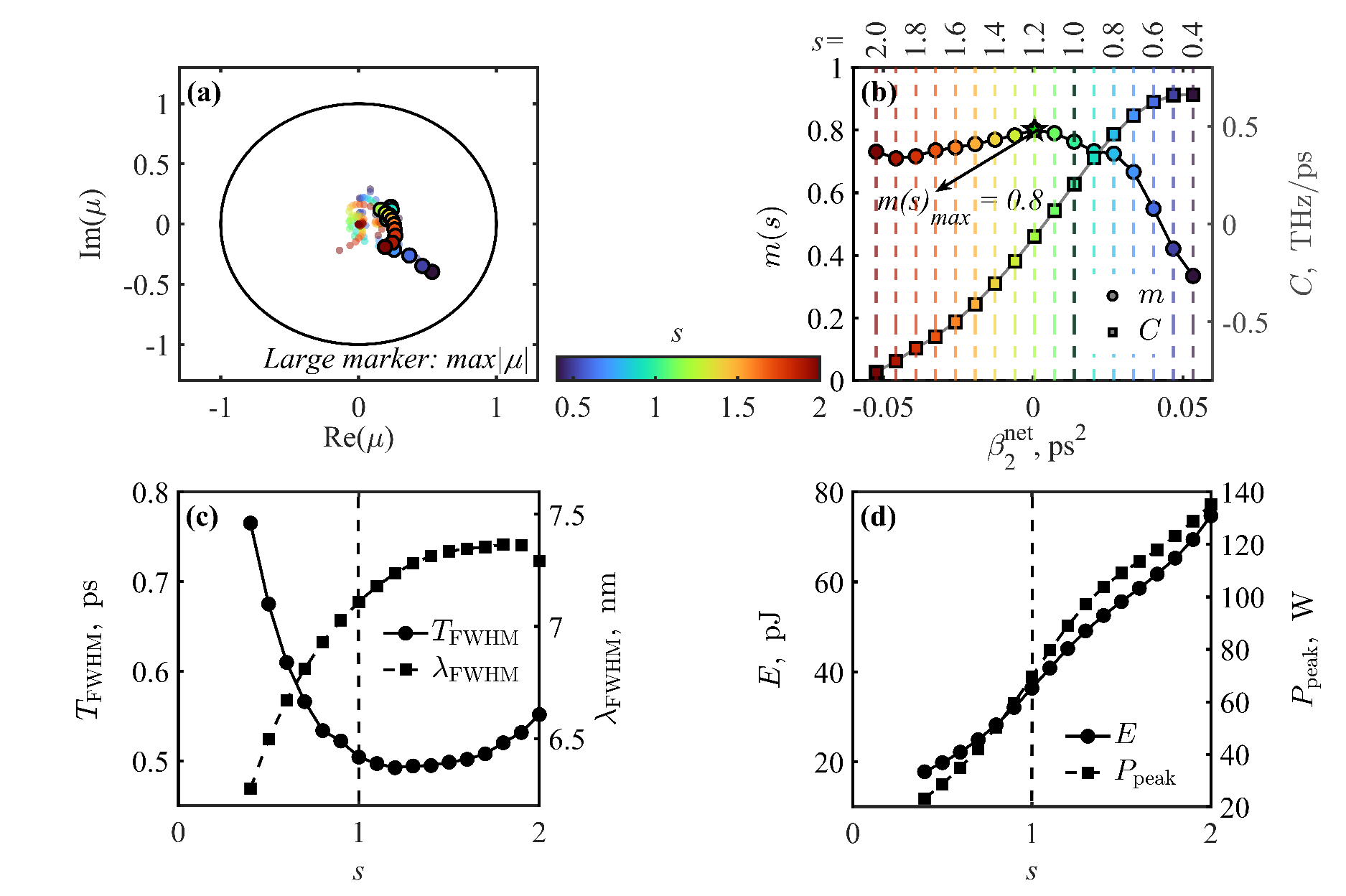} 
\caption{\textbf{Soliton stabilization behavior revealed through parametric sweep of anomalous fiber length.}
\textbf{(a)} Eigenvalue spectra for $s \in [0.4, 2]$ (colors indicate $s$ via colorbar). Large markers trace the dominant eigenvalue; small markers show all others clustered near the origin. Unit circle marks the stability boundary.
\textbf{(b)} Stability margin $m(s) = 1 - \rho_{\mathbf{J}}(s)$ (circles, left) and chirp coefficient $C$ (squares, right) versus net cavity dispersion $\beta_2^{\text{net}}$ and corresponding $s$. Maximum stability ($m \approx 0.81$) occurs at $\beta_2^{\text{net}} \approx 0$ ps$^2$, coinciding with a slightly negative chirp coefficient close to zero.
\textbf{(c,d)} Pulse evolution demonstrates nonlinear attractor dynamics: for $s > 1$, \textbf{(c)} pulse dimensions (temporal and spectral FWHM) remain constrained despite \textbf{(d)} considerable energy and power increases, confirming the soliton-shaping effect; for $s < 1$, pulse widths vary significantly with minimal energy change, indicating a transition to a dissipation dominated regime.}

\label{fig:pf2-robustness} 
\end{figure*}
To implement this linear analysis, we must construct the Jacobian matrix numerically. The matrix element $J_{ij}$ represents the sensitivity of the $i^{\mathrm{th}}$ output component to perturbations in the $j^{\mathrm{th}}$ input component. We employ central finite differences to compute these derivatives. For each column $j$, we perturb the  $j^{\mathrm{th}}$ component of $A^{\mathrm{st}}$ by $\pm\epsilon$ and evaluate:
\begin{equation}
J_{ij} = \frac{\mathcal{F}_{i}(A^{\mathrm{st}} + \epsilon e_{j}) - \mathcal{F}_{i}(A^{\mathrm{st}} - \epsilon e_{j})}{2\epsilon},
\end{equation}
where $e_{j}$ is the $j^{\mathrm{th}}$ canonical basis vector (unity at position $j$, zero elsewhere) and $\epsilon = 10^{-6} \max_{k}|A^{\mathrm{st}}_{k}|$ is the perturbation amplitude. By systematically perturbing along all $N$ basis directions, we construct the full Jacobian matrix. This central difference scheme achieves second-order accuracy $\mathcal{O}(\epsilon^2)$ and requires $2N$ round-trip evaluations. Once $\mathbf{J}$ is computed, stability is determined by examining its eigenvalue spectrum. The fixed point is asymptotically stable if all eigenvalues $\{\mu_i\}$ of $\mathbf{J}$ satisfy the spectral radius condition $\rho_{\mathbf{J}} = \max_i |\mu_i| < 1$, or equivalently, the stability margin $m = 1-\rho_{\mathbf{J}}>0$.

Applying this analysis to both the soliton-similariton and all-normal dispersion fiber laser configurations reveals a striking contrast (FIG.~\ref{fig:stability_analysis}). 
For the soliton-similariton laser with anomalous-dispersion passive fiber, all eigenvalues lie strictly inside the unit circle with $\rho_{\mathbf{J}} \approx 0.24$, yielding $m \approx 0.76$ (FIG.~\ref{fig:stability_analysis}(a,b)), which confirms that the system is linearly stable, as perturbation would decay eventually. In sharp contrast, the all-normal variant exhibits multiple eigenvalues $|\mu_i| > 1$ (FIG.~\ref{fig:stability_analysis}(c,d)), indicating that the system is unstable as the perturbation would grow with time. Linear stability analysis thus establishes that the anomalous-dispersion segment is the critical stabilizing element of the soliton-similariton laser, as there is no difference between the two systems except for this element.

The stabilizing effect of anomalous-dispersion fiber on the cavity is further illustrated by examining how the stability depends on its length. We define a dimensionless scaling factor $s \equiv L_3/L_{3,\text{base}}$, where $L_{3,\text{base}} = 2.9$ m is the initial PF2 length considered for Jacobian analysis above, and vary $s$ over the range $[0.4, 2]$, corresponding to nearly a five-fold variation in fiber length. For each value of $s$, we first obtain the steady-state fixed point $A^{\mathrm{st}}(s)$ by iterating the cavity map from the previously converged solution at $L_{3,\text{base}} = 2.9$ m, then construct the Jacobian matrix via numerical perturbations to extract the eigenspectrum $\{\mu_i(s)\}$, spectral radius $\rho_{\mathbf{J}}(s) = \max_i |\mu_i(s)|$, and stability margin $m(s) = 1 - \rho_{\mathbf{J}}(s)$. Additionally, we characterize output pulse properties for each $s$, including spectral and temporal widths, energy and peak power, and the linear chirp coefficient $C$, to identify correlations with stability behavior.

FIG.~\ref{fig:pf2-robustness} summarizes the result of this analysis across four complementary views. FIG.~\ref{fig:pf2-robustness}(a) displays eigenvalue evolution in the complex plane. The dominant eigenvalue (large markers) exhibits non-monotonic radial motion with respect to the unit circle, which defines the stability boundary. As the scaling factor $s$ increases, it moves inward from near the boundary at $s=0.4$ ($|\mu_{\max}|\approx 0.66$), approaches the origin at $s\approx 1.2$ ($|\mu_{\max}|\approx 0.24$ corresponding to maximum eigenvalue confinement, and then shows minimal variation for $s>1.2$. This dominant eigenvalue governs stability across the entire parameter range.

FIG.~\ref{fig:pf2-robustness}(b) plots the stability margin $m(s)$ (circles, left axis) and chirp coefficient $C(s)$ (squares, right axis) versus net cavity dispersion $\beta_2^{\mathrm{net}} = \left( L_1 \beta_{2,\mathrm{GF}} + L_2 \beta_{2,\mathrm{PF1}} + L_3 \beta_{2,\mathrm{PF2}} \right)$, with markers colored by $s$ to enable direct correlation with panel (a). The stability margin exhibits asymmetric behavior about $s = 1$. For $s < 1$ (decreasing anomalous fiber length), $m(s)$ decreases sharply from 0.76 at $s = 1$ to 0.34 at $s = 0.4$ as net dispersion becomes increasingly positive ($\beta_2^{\mathrm{net}} \rightarrow +0.055~\mathrm{ps}^2$)
. This regime is normal dispersion-dominated; therefore, a shortened anomalous segment cannot support soliton formation, resulting in a positive chirp (green squares) that increases with smaller $s$. Conversely, for $s > 1$ increasing anomalous fiber length, $m(s)$ rises from 0.76 to a maximum of 0.81 at $s \approx 1.2$ where $\beta_2^{\mathrm{net}} \approx 0$ ps$^2$. As  $s$ increases further, $\beta_2^{\mathrm{net}}$ becomes more negative and  $m(s)$ decreases only slightly, hovering around a value of 0.75, forming a broad stability plateau. This is also consistent with the results known for stretched-pulse fiber lasers, which indicate that timing jitter is minimized at near-zero dispersion \cite{Song2011OL_JitterRegimes}, a point that should correspond to maximum stability.  The chirp coefficient crosses zero in this maximum-stability region, then becomes increasingly negative for $\beta_2^{\mathrm{net}}< -0.01$ ps$^2$, indicating sufficient compression and soliton formation due to larger anomalous dispersion. 

FIG.~\ref{fig:pf2-robustness}(c,d) reveals distinct behavior in the two regimes. For $s \geq 1$, the temporal and spectral FWHM vary only marginally, despite the pulse energy doubling from 36 to 76 pJ and the peak power increasing from 69 to 135 W. This saturation demonstrates that pulse shape is locked by collective cavity dynamics and the attractor structure remains essentially unchanged even as total energy increases to compensate for longer propagation. For $s < 1$, temporal FWHM increases sharply from 0.50 to 0.75 ps while spectral FWHM decreases from 7.20 to 6.1 nm as $s$ decreases to 0.4, accompanied by energy dropping to 18 pJ and peak power to 23 W. This reduced soliton compression, indicated by temporal broadening and spectral narrowing, reflects insufficient anomalous dispersion for an effective soliton-shaping effect.
Thus, this section establishes that the anomalous-dispersion passive fiber provides robust linear stability to the soliton-similariton laser, with all eigenvalues confined well within the unit circle, while the all-normal variant exhibits instability. The results obtained establish that the anomalous segment clearly contributes to increasing the stability and robustness of the soliton-similariton fiber laser.

\section{Quantum-Limited Noise Properties of Soliton-Similariton Fiber Lasers}
\label{sec:citeref}

Linear stability guarantees that small deterministic perturbations decay, but noise from amplified spontaneous emission (ASE) continuously injects random fluctuations into the system at every round trip. A linearly stable system could, in principle, still accumulate noise if it lacks mechanisms to suppress these stochastic perturbations actively. The critical question is whether linear stability directly translates into noise suppression and whether the stability margin quantitatively predicts noise performance.
This section answers both these questions and quantifies the quantum-limited noise properties by comparing the soliton-similariton laser (System 1) with its all-normal variant (System 2), the same two-cavity comparison used in previous sections to isolate the role of the anomalous fiber.

We employ the numerical model developed by Paschotta \cite{Paschotta2004_APB_NoiseI,Paschotta2004_APB_NoiseII,Paschotta2010_OE_TimingPhaseNoise} for studying quantum-limited noise in mode-locked lasers, following procedures established for dissipative soliton systems \cite{Shin2015_OE_TimingJitterANDi,Zou2025_OE_PureQuarticQuantumDiffusion}. Amplified spontaneous emission (ASE) noise is modeled by adding a stochastic perturbation term $S_{\text{pert}}(z,t)$ to the field propagation in the gain fiber. In the frequency domain, this noise source exhibits the correlation \cite{Haus1991_JOSAB_QuantumNoiseSolitonRepeater}
\begin{align}
\left\langle \tilde{S}(z,\omega)\tilde{S}(z',\omega') \right\rangle &= \frac{1}{2\pi}\frac{1}{\left(\frac{\omega-\omega_0}{\Delta\omega}\right)^2+1}\,\theta\,\hbar\omega\Bigl(\exp(gh)-1\Bigr)h\nonumber\\
&\quad \times \delta(z-z')\delta(\omega-\omega'),
\label{eq:ASE_correlation}
\end{align}
where $\theta=3$ accounts for incomplete inversion, $\hbar$ is the Planck constant, $h$ is the gain fiber length, and the delta functions ensure frequency components remain uncorrelated, properly representing white-noise character of spontaneous emission.

Timing jitter and relative intensity noise (RIN) are computed following established methods \cite{Zou2025_OE_PureQuarticQuantumDiffusion,Shin2015_OE_TimingJitterANDi}. With ASE noise injected at each round trip according to Eq.~\eqref{eq:ASE_correlation}, the system is propagated for 10,000 round trips from the converged steady state. Timing deviations and pulse energy fluctuations relative to a noise-free reference trajectory yield the single-sideband power spectral densities $S_{\Delta t}(f)$ and $S_{\mathrm{RIN}}(f)$, which are integrated over the measurement bandwidth to obtain integrated timing jitter and integrated RIN. Integrated timing jitter and RIN are obtained by numerically integrating the respective PSDs. 

\begin{figure}[tbh]
\centering
\includegraphics[width=1.02\linewidth]{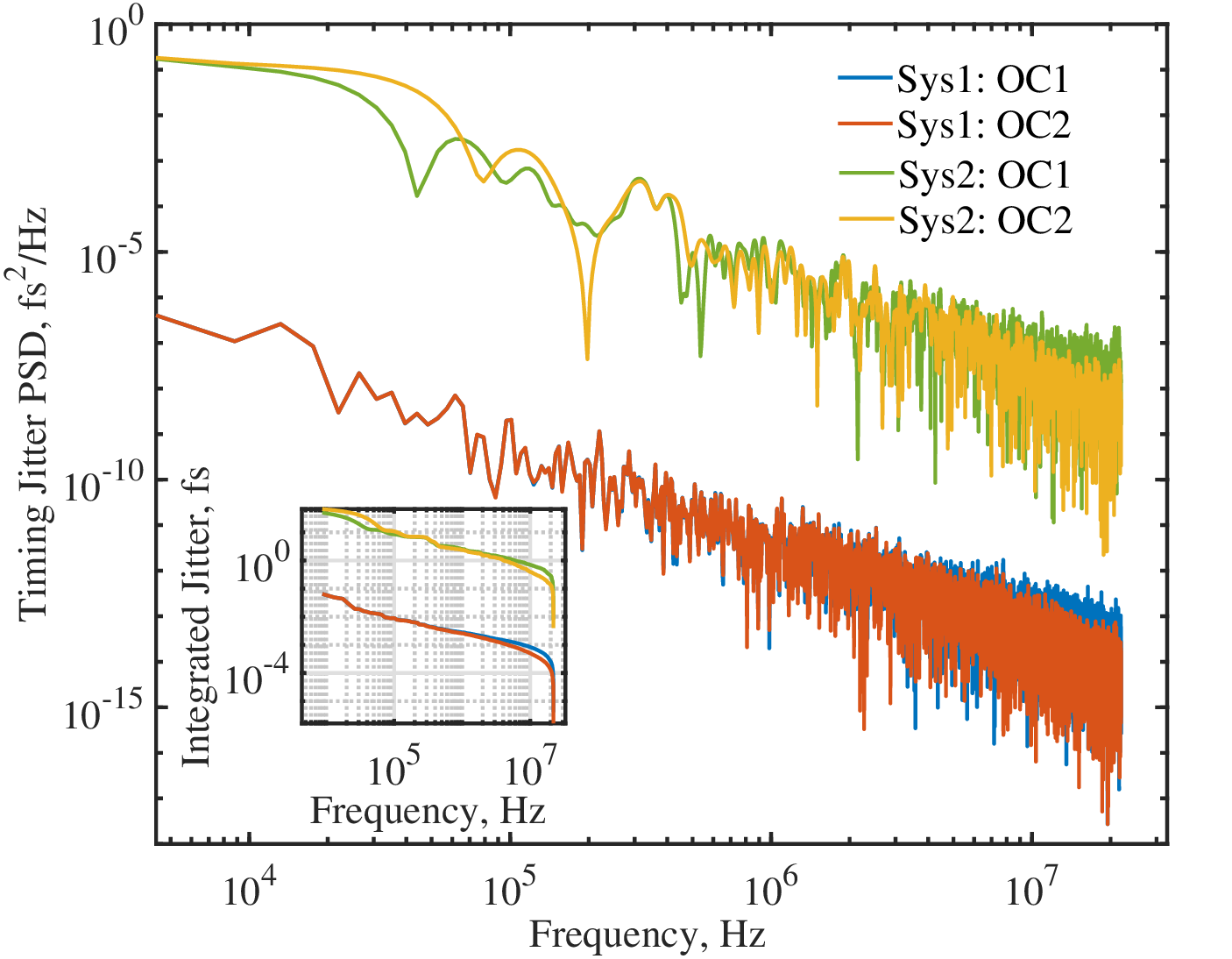}
\caption{\textbf{Timing jitter comparison between soliton-similariton (System~1) and all-normal (System~2) lasers at $s=1$.} Main panel: Single-sideband timing jitter PSD $S_{\Delta t}(f)$ for System~1 at OC1 and OC2 (blue/orange), and System~2 at OC1 and OC2 (green/yellow), plotted on log-log axes. System~1 maintains values near $10^{-6}\,\mathrm{fs}^2\,\mathrm{Hz}^{-1}$, while System~2 approaches $10^{-1}\,\mathrm{fs}^2\,\mathrm{Hz}^{-1}$, five orders of magnitude higher. Inset:  System~1 achieves sub-femtosecond integrated timing jitter, while System~2 exceeds 40\,fs.}
\label{fig:timing-jitter-psd}
\end{figure}

FIG.~\ref{fig:timing-jitter-psd} compares the single-sideband timing jitter PSD of both systems at their two output ports (OC1 and OC2). System~1 (soliton-similariton) maintains timing jitter PSD values at or below \SI{e-6}{\femto\second\squared\per\hertz} across the measurement bandwidth, while System~2 (all-normal) exhibits values approaching \SI{e-1}{\femto\second\squared\per\hertz}, approximately five orders of magnitude higher. This difference directly translates to the integrated timing jitter (inset): System~1 achieves sub-femtosecond stability ($<$\SI{1}{\femto\second}), while System~2 exceeds \SI{10}{\femto\second}.
Notably, the output from both couplers OC1 and OC2, for System 1 exhibits nearly identical timing jitter. Similar jitter values across different extraction points confirm that the anomalous segment provides effective noise filtering within the cavity, damping ASE-induced fluctuations before they can grow into macroscopic timing variations.

\begin{figure}[ht!]
\centering
\includegraphics[width=1.02\linewidth]{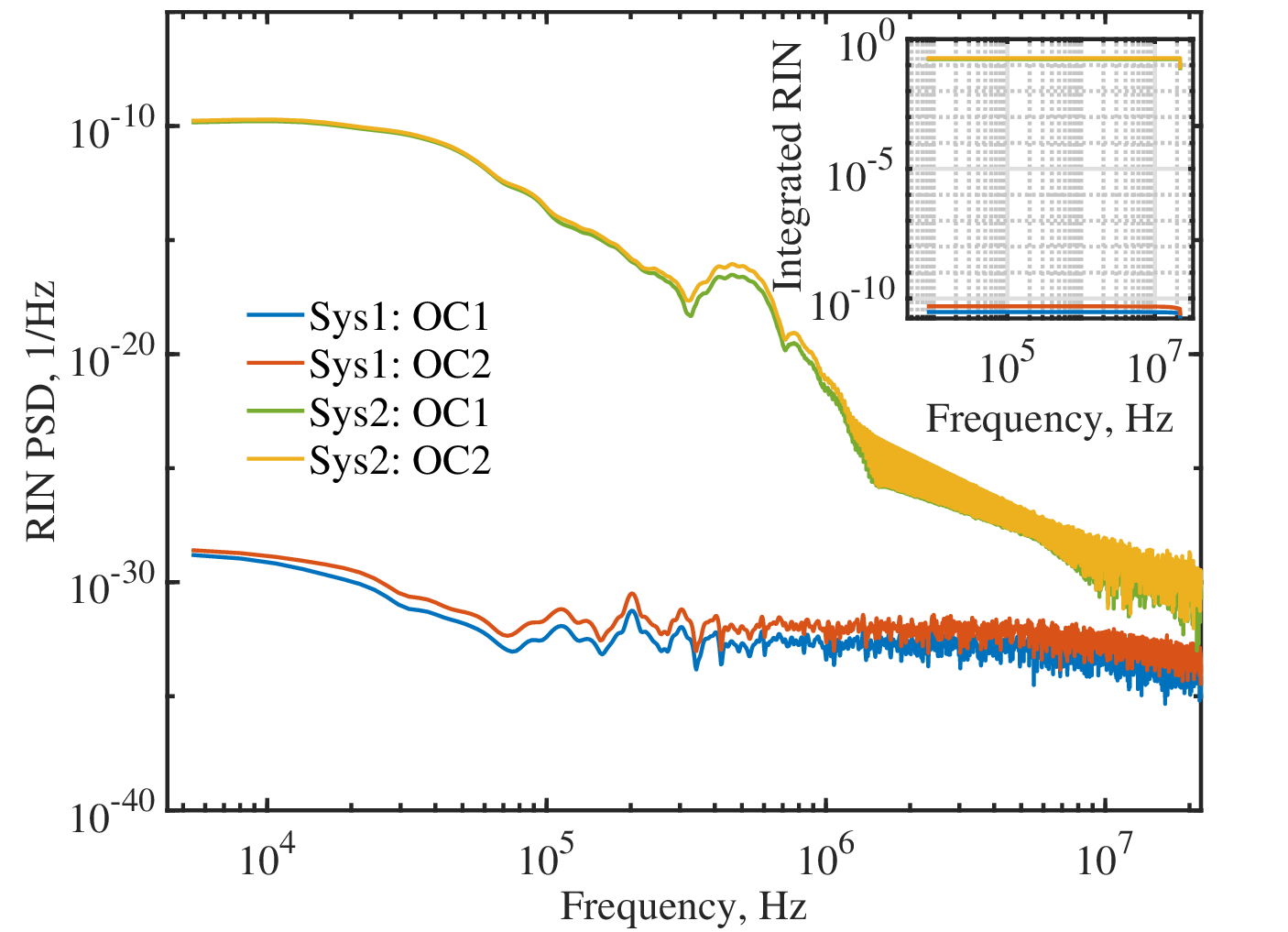}
\caption{\textbf{Relative intensity noise comparison between soliton-similariton (System~1) and all-normal (System~2) lasers at $s=1$.} Main panel: Single-sideband RIN PSD $S_I(f)$ for System~1 at OC1 and OC2 (blue/orange), and System~2 at OC1 and OC2 (green/yellow), plotted on log-log axes. System 1 exhibits RIN 10 orders of magnitude lower than System 2, with the integrated RIN rms (inset) showing a similar difference.}

\label{fig:rin-psd}
\end{figure}

FIG.~\ref{fig:rin-psd} presents the RIN comparison. Across the entire measurement range (\SIrange{10}{e5}{\hertz}), System~1 maintains RIN PSD at or below \SI{e-30}{\per\hertz}, while System~2 exhibits values between \SI{e-20}{\per\hertz} and \SI{e-10}{\per\hertz}, approximately ten orders of magnitude higher. The integrated RIN (inset) quantifies the cumulative effect: System~1 achieves values below \SI{e-10}{}, indicating fractional energy fluctuations of less than one part in $10^{10}$, while System~2 approaches unity, indicating severe amplitude instability.
System~1's absolute noise floor lies orders of magnitude below System~2 across the entire band.
The superior quantum-limited noise performance of the soliton-similariton laser can be understood through the interplay between soliton dynamics and noise suppression. In System~1, the anomalous-dispersion segment enables the formation of a soliton component, which is a nonlinear eigenstate of the propagation equation. Solitons possess a fundamental property: they maintain their shape through a precise balance between dispersion and nonlinearity, making them inherently resistant to perturbations. So when ASE-induced noise of the gain segment perturbs the pulse, the soliton dynamics in PF2 actively restore the pulse to its eigenstate, effectively filtering out timing and amplitude fluctuations that would otherwise accumulate. In contrast, System~2 lacks this restoring mechanism.  The all-normal-dispersion architecture supports only similariton (parabolic-pulse) dynamics in the gain fiber, which, while stable in the sense of convergence, does not yield the same noise-resilient eigenstate behavior. Perturbations accumulate over round trips without the self-correcting dynamics afforded by soliton formation, resulting in the observed five-to-ten order of magnitude degradation in noise performance. This physical mechanism also explains the linear stability analysis discussed in the previous section. Specifically, the eigenvalue confinement provided by the soliton attractor manifests as a stability margin  $m(s) = 1 - \rho_{\mathbf{J}}(s)$, where stronger confinement corresponds to more effective damping of perturbation.

\begin{figure}[tbh]
\centering
\includegraphics[width=1.0\linewidth]{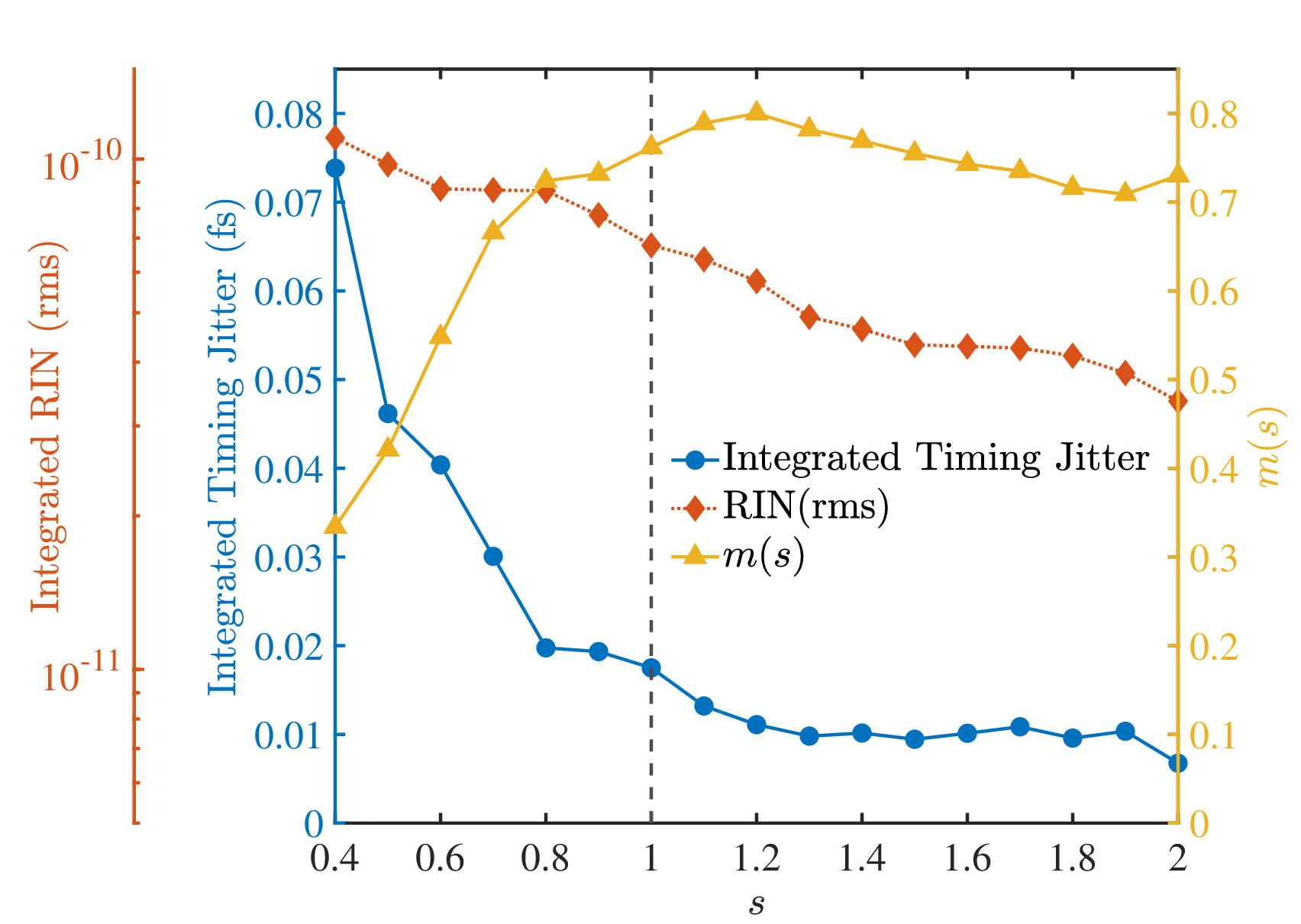}
\caption{\textbf{Quantum-limited noise metrics and stability margin versus anomalous fiber length.} Integrated timing jitter (solid circles, left axis), $\mathrm{RIN(rms)}$ (dotted pentagons, far-left axis), and stability margin $m(s)$ (triangles, right axis) as functions of the normalized anomalous fiber length $s$. The dashed vertical line marks the baseline length $s = 1$. The timing jitter and stability margin are strongly anti-correlated, with jitter decreasing as $m(s)$ increases, confirming that enhanced linear stability suppresses quantum-limited timing fluctuations. The $\mathrm{RIN(rms)}$ decreases monotonically with $s$ across the entire range.}
\label{fig:timing_jitter_vs_s}
\end{figure}

FIG.~\ref{fig:timing_jitter_vs_s} demonstrates the direct connection between stability margin and quantum-limited noise performance for System 1 and answers the critical question raised earlier: whether the stability margin quantitatively predicts noise performance. The stability-margin profile as shown in [FIG.~\ref{fig:pf2-robustness}(b)] is presented here along with integrated timing jitter and integrated RIN  for each value of $s$. The integrated timing jitter and RIN for each $s$ are calculated using the same procedure discussed above and are shown for frequencies up to 28 kHz. It can be observed that as $s$ increases above 1, the timing jitter decreases and remains nearly constant, mirroring the variation of the stability margin.  Conversely, as $s$ is reduced below 1, the timing jitter increases rapidly, consistent with the corresponding decrease in stability margin. They are anti-correlated but serve the same purpose. The integrated RIN, on the other hand, decreases monotonically with increasing $s$, confirming that amplitude fluctuations also improve with $ s$. This behavior reinforces the generality of the stability margin as a unified predictor of quantum-limited noise performance. Such a low integrated RIN in system 1 could be attributed to two restoring mechanisms: gain saturation and anomalous fiber length. Thus, the linear stability analysis prediction is validated by the quantum-limited noise model of the mode-locked laser, which also confirms that the soliton-shaping effect indeed contributes to the laser's improved noise performance. 

The observed anticorrelation between the stability margin and the integrated timing jitter clearly shows that the linear stability analysis can be effectively used as a predictive framework for the noise characteristics of mode-locked fiber lasers. This is important because estimating the stability margin requires significantly lower computational overhead. In more concrete terms, the estimation of the timing jitter for the soliton similariton (or the all-normal dispersion) laser required 21,000 round trip simulations. In contrast, the stability margin was estimated from a mere 2,000 round trips. The linear stability analysis thus provides a complementary framework with strong physical underpinnings for rapid noise optimization of mode-locked fiber lasers.

\section{Discussion and Outlook}
This work investigated the mechanisms underlying the exceptional stability and noise performance of soliton-similariton mode-locked fiber lasers by comparing them with an all-normal-dispersion variant. The all-normal cavity was constructed by replacing the soliton-supporting anomalous segment with a normally dispersive fiber. The basin of attraction analysis showed that both configurations converged to stable fixed points from arbitrary initial conditions. However, linear stability analysis of their fixed points via Jacobian eigenvalue decomposition revealed that the soliton-similariton cavity exhibited spectral radius $\rho < 1$, confirming asymptotic stability. In contrast, the all-normal variant showed $\rho > 1$, indicating intrinsic instability. This stability difference is directly manifested in the quantum-limited noise performance of the two cavities. Quantum-limited noise simulations showed soliton-similariton configuration achieved orders of magnitude lower timing jitter and RIN than the all-normal counterpart. The sub-femtosecond integrated timing jitter and sub-$10^{-10}$ fractional intensity noise positioned this laser among the quietest mode-locked fiber lasers at 1550~nm, reaching performance levels typically associated with solid-state systems.

These findings establish that the anomalous-dispersion passive fiber is not merely a pulse-shaping element but the critical component that provides superior quantum-limited noise performance through the soliton-shaping effect. Since quantum-limited timing jitter originates from amplified spontaneous emission (ASE) noise in the gain fiber, a noise source common to both cavity architectures, the observed orders-of-magnitude difference in noise performance arise from how each system controls the build-up of this unavoidable quantum noise. Soliton shaping in the anomalous segment possesses inherent restoring properties that not only damp ASE-induced fluctuations before they can grow into macroscopic timing variations but also underpin the asymptotic stability ($\rho < 1$) observed in our linear stability analysis of the soliton-similariton fiber laser. All-normal dispersion cavities, lacking this restoring force, fail to damp ASE quantum noise, resulting in noise performance orders of magnitude worse than that of soliton-similariton fiber lasers, which agrees with the observed linearly unstable characteristic ($\rho > 1$)  for the all-normal fiber laser. Thus, there exists a strong correlation between spectral radius and quantum-limited noise performance. Since the stability margin is directly related to the spectral radius, we introduced it as an alternative predictive framework for assessing the noise performance of fiber lasers. The determination of the stability margin is also highly computationally efficient, requiring at least an order of magnitude fewer round trips for assessing noise performance. These results indicate the possibility of establishing a quantitative link between the stability margin and associated measures with experimentally measurable quantities like timing jitter and intensity noise, which will be explored in future work.

In summary, this comprehensive stability and noise analysis establishes the soliton-similariton fiber laser as a fundamentally robust architecture, with the anomalous-dispersion passive fiber serving as the critical element that simultaneously ensures linear stability and low timing jitter through the soliton-shaping effect. Remarkably, this stabilization improves with increasing anomalous fiber length across a wide range of values, positioning soliton-similariton lasers as premier platforms for precision photonic applications that demand both long-term
stability and low timing jitter. Additionally, the reported results establish the Jacobian eigenvalue-based linear stability analysis as a highly effective approach for studying laser stability and robustness, paving the way for novel design approaches for fiber lasers and lasers in general.

\begin{acknowledgments}
\vspace{-0.25em}
The authors acknowledge funding from the DST-CRG Grant, awarded by the Science and Engineering Research Board, now known as Anusandhan National Research Foundation, a statutory body of the Department of Science and Technology (DST), Government of India.
\vspace{-0.25em}
\end{acknowledgments}

\appendix

\nocite{*}

\bibliography{Manuscript}% Produces the bibliography via BibTeX.

@article{Anderson1993_JOSAB_WBFree,
  author  = {Anderson, D. and Desaix, M. and Karlsson, M. and Lisak, M. and Quiroga-Teixeiro, M. L.},
  journal = {J. Opt. Soc. Am. B},
  year    = {1993},
  volume  = {10},
  number  = {7},
  pages   = {1185--1190},
  doi     = {10.1364/JOSAB.10.001185}
}

@article{Fermann2000_PRL_SelfSimilarPRL,
  author  = {Fermann, M. E. and Kruglov, V. I. and Thomsen, B. C. and Dudley, J. M. and Harvey, J. D.},
  journal = {Phys. Rev. Lett.},
  year    = {2000},
  volume  = {84},
  number  = {26},
  pages   = {6010--6013},
  doi     = {10.1103/PhysRevLett.84.6010}
}

@article{Ilday2004_PRL_SelfSimilarLaser,
  author  = {Ilday, F. {\"O}. and Buckley, J. R. and Clark, W. G. and Wise, F. W.},
  journal = {Phys. Rev. Lett.},
  year    = {2004},
  volume  = {92},
  number  = {21},
  pages   = {213902},
  doi     = {10.1103/PhysRevLett.92.213902}
}

@article{Renninger2010_PRA_ANDiSelfSimilar,
  author  = {Renninger, W. H. and Chong, A. and Wise, F. W.},
  journal = {Phys. Rev. A},
  year    = {2010},
  volume  = {82},
  number  = {2},
  pages   = {021805},
  doi     = {10.1103/PhysRevA.82.021805}
}

@article{Chong2007_OL_ANDi20nJ,
  author  = {Chong, A. and Renninger, W. H. and Wise, F. W.},
  journal = {Opt. Lett.},
  year    = {2007},
  volume  = {32},
  number  = {16},
  pages   = {2408--2410},
  doi     = {10.1364/OL.32.002408}
}

@article{Chong2015_RPP_SimilaritonReview,
  author  = {Chong, Andy and Wright, Logan G. and Wise, Frank W.},
  journal = {Rep. Prog. Phys.},
  year    = {2015},
  volume  = {78},
  number  = {11},
  pages   = {113901},
  doi     = {10.1088/0034-4885/78/11/113901}
}

@article{Oktem2010_NatPhoton_SSFL,
  author  = {Oktem, B{\"u}lent and {\"U}lg{\"u}d{\"u}r, Co{\c{s}}kun and Ilday, F. {\"O}mer},
  journal = {Nat. Photonics},
  year    = {2010},
  volume  = {4},
  number  = {5},
  pages   = {307--311},
  doi     = {10.1038/nphoton.2010.33}
}

@article{Zhang2010_OL_SSFL_BirefringentFilter,
  author  = {Zhang, H. and Oktem, B. and Ilday, F. {\"O}.},
  journal = {Opt. Lett.},
  year    = {2010},
  volume  = {35},
  number  = {20},
  pages   = {3252--3254},
  doi     = {10.1364/OL.35.003252}
}

@article{Bale2010_OL_StrongSpectralFiltering,
  author  = {Bale, B. G. and Wabnitz, S.},
  journal = {Opt. Lett.},
  year    = {2010},
  volume  = {35},
  number  = {14},
  pages   = {2466--2468},
  doi     = {10.1364/OL.35.002466}
}

@article{Wang2017_JOSAB_SpectralFilteringTransition,
  author  = {Wang, Zhiqiang and Zhan, Li and Fang, Xiao and Luo, Hao},
  journal = {J. Opt. Soc. Am. B},
  year    = {2017},
  volume  = {34},
  number  = {11},
  pages   = {2325--2333},
  doi     = {10.1364/JOSAB.34.002325}
}

@article{Lapre2019_SciRep_RealtimeInstabilities,
  author  = {Lapre, C. and Billet, C. and Meng, F. and Ryczkowski, P. and Sylvestre, T. and Finot, C. and Genty, G. and Dudley, J. M.},
  journal = {Sci. Rep.},
  year    = {2019},
  volume  = {9},
  pages   = {13950},
  doi     = {10.1038/s41598-019-50022-5}
}

@article{meng2020instabilities,
  author={Meng, Fanchao and Lapre, Coraline and Billet, Cyril and Genty, Go{\"e}ry and Dudley, John M},
  journal={Opt. Lett.},
  volume={45},
  number={5},
  pages={1232--1235},
  year={2020},
  doi     = {https://doi.org/10.1364/OL.386110}
}

@article{Mohamed2024_NatCommun_EMS,
  author  = {Mohamed, Mostafa I. and Coillet, Aur{\'e}lien and Grelu, Philippe},
  journal = {Nat. Commun.},
  year    = {2024},
  volume  = {15},
  number  = {1},
  pages   = {8875},
  doi     = {10.1038/s41467-024-52954-7}
}

@article{Paschotta2004_APB_NoiseI,
  author  = {Paschotta, R{\"u}diger},
  journal = {Appl. Phys. B},
  year    = {2004},
  volume  = {79},
  number  = {2},
  pages   = {153--162},
  doi     = {10.1007/s00340-004-1547-x}
}

@article{Paschotta2004_APB_NoiseII,
  author  = {Paschotta, R{\"u}diger},
  journal = {Appl. Phys. B},
  year    = {2004},
  volume  = {79},
  number  = {2},
  pages   = {163--173},
  doi     = {10.1007/s00340-004-1548-9}
}

@article{Grelu2012_NatPhoton_DissipativeSolitons,
  author  = {Grelu, Philippe and Akhmediev, Nail},
  journal = {Nat. Photonics},
  year    = {2012},
  volume  = {6},
  number  = {2},
  pages   = {84--92},
  doi     = {10.1038/nphoton.2011.345}
}

@article{Paschotta2010_OE_TimingPhaseNoise,
  author  = {Paschotta, R{\"u}diger},
  journal = {Opt. Express},
  year    = {2010},
  volume  = {18},
  number  = {5},
  pages   = {5041--5054},
  doi     = {10.1364/OE.18.005041}
}

@article{Shin2015_OE_TimingJitterANDi,
  author  = {Shin, Junho and Jung, Kwangyun and Song, Youjian and Kim, Jungwon},
  journal = {Opt. Express},
  year    = {2015},
  volume  = {23},
  number  = {17},
  pages   = {22898--22906},
  doi     = {10.1364/OE.23.022898}
}

@article{Zou2025_OE_PureQuarticQuantumDiffusion,
  author  = {Zou, Defeng and Guo, Penglai and Liu, Runmin and Zhang, Aoyan and Li, Jialong and Chen, Gina Jinna and Dang, Hong and Li, Xiaohui and Song, Youjian and Shum, Perry Ping},
  journal = {Opt. Express},
  year    = {2025},
  volume  = {33},
  number  = {1},
  pages   = {1437--1447},
  doi     = {10.1364/OE.545988}
}

@article{Haus1991_JOSAB_QuantumNoiseSolitonRepeater,
  author  = {Haus, Hermann A.},
  journal = {J. Opt. Soc. Am. B},
  year    = {1991},
  volume  = {8},
  number  = {5},
  pages   = {1122--1126},
  doi     = {10.1364/JOSAB.8.001122}
}

@book{AGRAWAL817042,
    author ={Govind P. Agrawal} ,
    title = {Nonlinear Fiber Optics },
    publisher = {Academic Press, Cambridge},
    year = {2021},
    pages = {46-50},
    url={https://www.sciencedirect.com/science/article/abs/pii/B9780128170427000117},
}

@article{Renninger2012JSTQE,
  author  = {Renninger, William H. and Chong, Andy and Wise, Frank W.},
  journal = {IEEE J. Sel. Top. Quantum Electron.},
  volume  = {18},
  number  = {1},
  pages   = {389--398},
  year    = {2012},
  month   = jan,
  doi     = {10.1109/JSTQE.2011.2157462}
}

@article{Yarutkina2013OE,
  author  = {Yarutkina, I. A. and Shtyrina, O. V. and Fedoruk, M. P. and Turitsyn, S. K.},
  journal = {Opt. Express},
  volume  = {21},
  number  = {10},
  pages   = {12942--12950},
  year    = {2013},
  month   = may,
  doi     = {10.1364/OE.21.012942}
}

@article{Kapitula2002,
  author  = {Kapitula, T. and Kutz, J. N. and Sandstede, B.},
  journal = {J. Opt. Soc. Am. B},
  volume  = {19},
  pages   = {740--746},
  year    = {2002},
  doi     = {10.1364/JOSAB.19.000740}
}

@article{Shinglot2025,
  author  = {Shinglot, V. and Zweck, J.},
  journal = {SIAM J. Appl. Math.},
  volume  = {85},
  pages   = {420--449},
  year    = {2025},
  doi     = {10.1137/23M1598106}
}

@article{schreiber2007study,
  author  = {Schreiber, Thomas and Orta{\c{c}}, B{\"u}lend and Limpert, Jens and T{\"u}nnermann, Andreas},
  journal = {Opt. Express},
  volume  = {15},
  number  = {13},
  pages   = {8252--8262},
  year    = {2007},
  doi     = {10.1364/OE.15.008252},
  url     = {https://opg.optica.org/fulltext.cfm?uri=oe-15-13-8252}
}

@article{Haus1995StretchedPulseAPM,
  author  = {Haus, H. A. and Tamura, K. and Nelson, L. E. and Ippen, E. P.},
  journal = {IEEE J. Quantum Electron.},
  volume  = {31},
  number  = {3},
  pages   = {591--598},
  year    = {1995},
  doi     = {10.1109/3.364417}
}

@article{HausIslam1985JQE,
  author  = {Haus, H. A. and Islam, M. N.},
  journal = {IEEE J. Quantum Electron.},
  volume  = {21},
  number  = {8},
  pages   = {1172--1188},
  year    = {1985},
  doi     = {10.1109/JQE.1985.1072805},
  url     = {https://doi.org/10.1109/JQE.1985.1072805}
}

@article{Kelly1992EL,
  author  = {Kelly, S. M. J.},
  journal = {Electron. Lett.},
  volume  = {28},
  number  = {8},
  pages   = {806--807},
  year    = {1992},
  doi     = {10.1049/el:19920508},
  url     = {https://doi.org/10.1049/el:19920508}
}

@article{HausMecozzi1993JQE,
  author  = {Haus, H. A. and Mecozzi, A.},
  journal = {IEEE J. Quantum Electron.},
  volume  = {29},
  number  = {3},
  pages   = {983--996},
  year    = {1993},
  doi     = {10.1109/3.206583},
  url     = {https://doi.org/10.1109/3.206583}
}

@article{NamikiYuHaus1996JOSAB,
  author  = {Namiki, S. and Yu, C. X. and Haus, H. A.},
  journal = {J. Opt. Soc. Am. B},
  volume  = {13},
  number  = {12},
  pages   = {2817--2824},
  year    = {1996},
  doi     = {10.1364/JOSAB.13.002817},
  url     = {https://doi.org/10.1364/JOSAB.13.002817}
}

@article{Tang2005PRA,
  author  = {Tang, D. Y. and Zhao, L. M. and Zhao, B. and Liu, A. Q.},
  journal = {Phys. Rev. A},
  volume  = {72},
  pages   = {043816},
  year    = {2005},
  doi     = {10.1103/PhysRevA.72.043816}
}

@article{Tamura1993OL,
  author  = {Tamura, K. and Ippen, E. P. and Haus, H. A. and Nelson, L. E.},
  journal = {Opt. Lett.},
  volume  = {18},
  number  = {13},
  pages   = {1080--1082},
  year    = {1993},
  doi     = {10.1364/OL.18.001080},
  url     = {https://doi.org/10.1364/OL.18.001080}
}

@article{Tamura1994APL,
  author  = {Tamura, K. and Nelson, L. E. and Haus, H. A. and Ippen, E. P.},
  journal = {Appl. Phys. Lett.},
  volume  = {64},
  number  = {2},
  pages   = {149--151},
  year    = {1994},
  doi     = {10.1063/1.111547},
  url     = {https://doi.org/10.1063/1.111547}
}

@article{Namiki1997JQE_NoiseSPFL_I,
  author  = {Namiki, S. and Haus, H. A.},
  journal = {IEEE J. Quantum Electron.},
  volume  = {33},
  number  = {5},
  pages   = {649--659},
  year    = {1997},
  doi     = {10.1109/3.572138}
}

@article{Song2011OL_JitterRegimes,
  author  = {Song, Youjian and Jung, Kwangyun and Kim, Jungwon},
  journal = {Opt. Lett.},
  volume  = {36},
  number  = {10},
  pages   = {1761--1763},
  year    = {2011},
  doi     = {10.1364/OL.36.001761}
}

@article{Tamura1995,
  author   = {Tamura, K. and Ippen, E. P. and Haus, H. A.},
  journal  = {Appl. Phys. Lett.},
  volume   = {67},
  pages    = {158--160},
  year     = {1995}
}

\end{document}